\documentclass[aps,prb,showpacs,twocolumn,floats,epsfig,pdflatex]{revtex4}
\usepackage{epsfig}
\usepackage{amsmath}
\usepackage{amsfonts}
\usepackage{graphicx}
\usepackage{amssymb}
\usepackage{amsbsy}
\usepackage{xcolor}
\usepackage{ulem}

\begin{document}

\title{Restoring coherence via aperiodic drives in a many-body quantum system}
\author{Bhaskar Mukherjee$^1$, Arnab Sen$^1$, Diptiman
Sen$^2$, and K. Sengupta$^1$} \affiliation{$^1$School of Physical
Sciences, Indian Association for the
Cultivation of Science, Kolkata 700032, India \\
$^2$Centre for High Energy Physics and Department of Physics, Indian
Institute of Science, Bengaluru 560012, India}
\date{\today}

\begin{abstract}

We study the unitary dynamics of randomly or quasi-periodically
driven tilted Bose-Hubbard (tBH) model in one dimension deep inside
its Mott phase starting from a $\mathbb{Z}_2$ symmetry-broken state.
The randomness is implemented via a telegraph noise protocol in the
drive period while the quasi-periodic drive is chosen to correspond
to a Thue-Morse sequence. The periodically driven tBH model (with a
square pulse protocol characterized by a time period $T$) is known
to exhibit transitions from dynamical regimes with long-time
coherent oscillations to those with rapid thermalization. Here we
show that starting from a regime where the periodic drive leads to
rapid thermalization, a random drive, which consists of a random
sequence of square pulses with period $T+\alpha dT$, where
$\alpha=\pm 1$ is a random number and $dT$ is the amplitude of the
noise, restores long-time coherent oscillations for special values
of $dT$. A similar phenomenon can be seen for a quasi-periodic drive
following a Thue-Morse sequence where such coherent behavior is
shown to occur for a larger number of points in the $(T, dT)$ plane
due to the additional structure of the drive protocol. We chart out
the dynamics of the system in the presence of such aperiodic drives,
provide a qualitative analytical understanding of this phenomenon,
point out the role of quantum scars behind it, and discuss
experiments which can test our theory.

\end{abstract}


\maketitle

\section{Introduction}

It is well-known that the bulk energy spectrum of any non-integrable
many-body quantum system satisfies the eigenstate thermalization
hypothesis (ETH)\cite{rev1,deustch1,srednicki1,rigol1}. ETH provides
a natural explanation of eventual thermalization starting from a
generic non-equilibrium many-body quantum state. One of the
consequence of ETH is the decay of coherent quantum oscillations in
the expectation value of a generic local operator during its
evolution as the system reaches a steady state
\cite{steady1,steady2}; such a decay is characterized by a
system-dependent timescale, $\tau_{th}$, which is identified as the
thermalization time. The divergence of the thermalization time
leading to failure of ETH is seen in many-body localized systems
where strong disorder leads to non-ergodicity \cite{mblref}. Another
weaker violation of ETH occurs in certain disorder-free systems due
to presence of special energy eigenstates, dubbed as many-body
quantum scars, in the spectrum of the bulk eigenstates of these
system \cite{scarqm1,scarref1,scarref2,scarref3}. The consequence of
presence of such states in the eigenstates of the Hamiltonian
describing a Rydberg chain was experimentally verified via
observation of long-time coherence oscillation of Rynderg
excitations \cite{scarref1}. It was noted that such long-time
oscillations, which occurs only if the starting state is
$|\mathbb{Z}_2\rangle$ (a state with one Rydberg excitation on every
alternate site), could not be explained within the framework of ETH
\cite{scarref2,scarref3}. Instead, their presence occurs due to the
existence of quantum scars which are states with finite energy
density but sub-thermal half-chain entanglement: $S_{L/2} \sim \ln
L$ where $L$ refers to the total number of sites in the chain. These
states have large and finite overlap with $|\mathbb{Z}_2\rangle$ and
form an almost closed subspace in the system's Hilbert space. The
evolution of the system, starting from the $|\mathbb{Z}_2\rangle$
state, therefore occurs within this almost closed subspace leading
to breakdown of ergodicity and failure of ETH.

More recently, the fate of such scar-induced coherent oscillations
were studied in the context of a periodically driven Rydberg chain
\cite{bm1}. It was shown that for high drive frequencies where the
properties of the system can be understood in terms of a Floquet
Hamiltonian $H_F$ \cite{fl1} computed using Magnus expansion
\cite{fl2}, the bulk eigenstates of $H_F$ host scars whose presence
lead to long-time coherent oscillations in the density-density
correlation function of the Rydberg atoms. In contrast, at low
frequencies, $H_F$ do not host scars and the correlation function
shows expected thermalization consistent with ETH prediction. In
between, at intermediate drive frequencies, the system undergoes
several reentrant transitions between thermal and coherent regimes.
The reason for such transition could be analytically, albeit
qualitatively, understood by noting that a special class of local
terms in $H_F$, which are responsible for hosting scars in its
eigenspectrum, have vanishing amplitude at special drive
frequencies. Near these drive frequencies, the system crosses over
from coherent to thermal behavior. The density-density correlator
displays increasingly shorter $\tau_{th}$ as these special
frequencies are approached. The fastest thermalization occurs in the
vicinity of these special frequencies where coherent oscillations
are almost absent~\cite{bm1}.

In this work, we study the driven tilted Bose-Hubbard model (tBH) in
the presence of random and quasiperiodic drives. The model
Hamiltonian we use for such a study involves a representation of
this model in terms of Ising spins \cite{bm1,subir1}and is given by
\begin{eqnarray}
H_0 &=& \sum_{j} \left(-w \tilde \sigma_j^x + \frac{\lambda}{2}
\sigma_j^z \right) \label{ham1}
\end{eqnarray}
where $\sigma_j^{\alpha}$ for $\alpha=x,y,z$ denote Pauli spin
matrices on site $j$ of the chain, $\tilde \sigma_j^{\alpha} =
P_{j-1} \sigma_j^{\alpha} P_{j+1}$, $P_j= (1-\sigma_j^z)/2$ is a
projection operator which projects to the $|\downarrow\rangle$
state, and $w$ and $\lambda$ denote strength of the effective
transverse and longitudinal field terms of the spin model.
Furthermore, there is an additional constraint that the spins on any
two neighboring sites cannot simultaneously be
$|\uparrow,\uparrow\rangle$.

In what follows, we shall always be in the regime
$w/|\lambda| \ll 1$ and drive $\lambda$ according to some given
protocol keeping $w$ fixed. More specifically, in this work, we
shall be studying two drive protocols. The first involves a random
sequences of square pulses with period $T_{\pm} = T+\alpha dT$,
where $\alpha=\pm 1$ is a random number and $dT$ is the strength of
the noise. The second protocol involve a quasiperiodic drive which
follows the Thue-Morse sequence for which the sequence of numbers
$\{\alpha_i\}$, rather than being random, satisfies
\begin{eqnarray}
\{\alpha_{2n}\} &=& \{\alpha_{n}\}, \quad
\{\alpha_{2n+1}\}=-\{\alpha_n\} \label{tms}
\end{eqnarray}
with $\alpha_0=-1$ \cite{tmref1,tmdyn1}. The drive period for the
$n^{\rm th}$ square pulse following the Thue-Morse protocol is then
given by $T_n= T+\alpha_n dT$.

The central results that we obtain from such a study are as follows.
First, starting from the initial state $|\psi_0\rangle =
(|\mathbb{Z}_2\rangle +|{\bar{\mathbb{Z}}_2}\rangle)/\sqrt{2}$
(where $|\bar{\mathbb{Z}}_2\rangle$ is the time-reversed counterpart
of $|\mathbb{Z}_2\rangle$ and $|\mathbb{Z}_2\rangle= |\uparrow
\downarrow \uparrow ...\rangle$), for the case of random protocol,
we show that the presence of the telegraphic noise with specific
noise strength $dT$ may restore coherent oscillations of the spin
correlation functions even when such correlators shows ETH predicted
thermalization in the absence of noise. We demonstrate this by exact
numerics on finite sized Rydberg chains with length $L\le 26$.
Second, using the fact that $w/|\lambda(t)| \ll 1$ at
all times, we provide an analytic explanation of this phenomenon.
Our results allow us to provide a phase diagram as a function of
$dT$ and $T$ which indicates the specific values of $dT$ and $T$ at
which we expect such coherent behavior; these results agree
qualitatively with the prediction of exact numerics. Moreover, our
analysis elucidate the role of quantum scars behind this phenomenon.
Third, we demonstrate the presence of coherence restoration for
dynamics using Thue-Morse sequence at specific values of $dT$ and
provide a semi-analytic explanation for their occurrence. Finally,
we discuss experiments involving ultracold Rydberg chain which can
test our theory.

The plan of the paper is as follows. In Sec.\ \ref{sec2}, we discuss
the model Hamiltonian and its relation to the Hamiltonian governing
the dynamics of 1D Rydberg atoms. This is followed by Secs.\
\ref{sec3} and \ref{sec4} where we present our results on random and
quasiperiodic drive protocols. Finally, we chart out our main
results, discuss experiments which can be used to verify them, and
conclude in Sec.\ \ref{sec5}.

\section{Models}
\label{sec2}

In this section, we chart out the model used in the present study
and its relation to Hamiltonian describing atoms in a ultracold
Rydberg chain. We start with the tilted Bose-Hubbard model given by
\begin{eqnarray}
H &=& -w_0 \sum_{\langle ij\rangle} (b_i^{\dagger} b_j + {\rm h.c.})
- \sum_i (\mu_0 + E_1 i) n^b_i  \nonumber\\
&& + \sum_i  \frac{U}{2} n^b_i (n^b_i-1) \label{bham1}
\end{eqnarray}
where $b_i$ ($b^\dagger_i$) denotes the boson annihilation
(creation) operator on site $i$ of a 1D chain, $n^b_i= b_i^{\dagger}
b_i$ is the number operator for bosons, $E_1$ denotes the effective
electric field for the bosons which controls the magnitude of the
tilt, $\mu_0$ is the boson chemical potential, $w_0$ is the
amplitude for nearest-neighbor hopping , and $U$ is the on-site
interaction strength.

It is well-known that the effective low-energy description of these
model can be achieved in terms of dipoles living on a link $\ell$
between two consecutive lattice sites $j$ and $j'$. The creation
operator for these dipoles can be written as $d_{\ell}^{\dagger} =
b_j^{\dagger} b_{j-1}/\sqrt{n_0(n_0+1)}$, where $n_0$ is the ground
state occupation of the parent Mott state without the tilt. In terms
of these dipoles the effective low-energy Hamiltonian can be written
as \cite{subir1}
\begin{eqnarray}
H_d &=& \sum_{\ell}\left( -w (d_{\ell} + d_{\ell}^{\dagger}) +
\lambda n_{\ell}^d \right) \label{diham1}
\end{eqnarray}
where $w= w_0 \sqrt{n_0(n_0+1)}$ is the amplitude for spontaneous
creation and annihilation of dipoles, $\lambda=(U-E_1)$ is the
dipole chemical potential, and $n_{\ell}^d= d_{\ell}^{\dagger}
d_{\ell}$ is the dipole number operator. The Hamiltonian $H_d$ is to
be supplemented by two constraint conditions that make it
non-integrable: $n_{\ell}^d \le 1$ and $n_{\ell}^d n^d_{\ell+1}=0$.
The first ensures that the maximum number of dipoles on any link is
unity and the second guarantees that there are no states with two
dipoles on neighboring links. For large positive $\lambda/w$, the
ground state of the model consists of a dipole vacuum while for
large negative $\lambda/w$, it is a $\mathbb{Z}_2$ symmetry broken
state with maximal number of dipoles which we denote as
$|\mathbb{Z}_2\rangle$. These two states are separated by a quantum
phase transition at $\lambda/w=-1.31 \sqrt{n_0(n_0+1)}$ which
belongs to the Ising universality class. The non-equilibrium
dynamics of the model, starting from the dipole vacuum or
$|\mathbb{Z}_2\rangle$ has been studied for quench, ramp and
periodic protocols \cite{dipoledyn}. The model has
also been experimentally realized using ultracold boson chains
\cite{dipoleexp1}.

In what follows, we shall use a spin representation of this dipole
model which allows us to implement the constraint in a easier
manner. To this end, we use the transformation $\sigma_{\ell}^{x[y]}
= [i](d_{\ell} +[-] d_{\ell}^{\dagger})$ and $\sigma_{\ell}^z =
2n_{\ell}^d-1$. In terms of the spin variables, one obtains
\begin{eqnarray}
H_s &=& \sum_{\ell} \left(-w \sigma_{\ell}^x + \lambda
\sigma_{\ell}^z/2 \right) \label{spinham1}
\end{eqnarray}
with the constraint $(1+\sigma_{\ell}^z)(1+ \sigma_{\ell+1}^z)=0$.
It was noted in Ref.\ \onlinecite{scarref3} that this constraint
condition could be implement by a local projection operator
$P_{\ell} =(1-\sigma_{\ell}^z)/2$ which enables one to equate $H_s$
to $H_0$. For $\lambda=0$, $H_0$ only contains a single term and has
been referred to as the PXP model \cite{scarref2,scarref3}. It is
well-known that for $\lambda=0$, the eigenspectrum of $H_0$ hosts
quantum scars and lead to long time coherent oscillation of
$O_{\ell_1 \ell_2} = \langle \sigma^z_{\ell_1}
\sigma^z_{\ell_1+\ell_2}\rangle$ \cite{scarref1}.

Such a long-time coherent oscillatory behavior of the spin
correlator was experimentally verified in a Rydberg chain. The
effective low-energy Hamiltonian for these Rydberg atoms can be
written as \cite{scarref1}
\begin{eqnarray}
H_{\rm Ryd} &=& \sum_i (\Omega \sigma_i^x + \Delta n_i) + \sum_{ij}
V_{ij} n_i n_j  \label{hryd}
\end{eqnarray}
where $n_i \le 1$ is the number of Rydberg atoms on site $i$,
$\Delta$ denotes the detuning parameter used to facilitate a Rydberg
excitation, $V_{ij} \sim 1/|i-j|^3$ is the interaction between them,
$\sigma_x^i= |R_i\rangle \langle G_i| +{\rm h.c.}$ describes
coupling between atoms in the Rydberg excited ($|R_i\rangle$) and
ground ($|G_i\rangle$) states. We note that experiments on these
system can tune $V_{ij}$ such that $V_{i i+1} \gg \Delta, \Omega \gg
V_{i i+2}$ \cite{scarref1}; in this case, the interaction acts as a
constraint of not having two Rydberg excitations on neighboring
sites. In this regime $H_{\rm RYD}$ can be directly mapped to $H_0$
with $\Omega \to -w$, $n_i \to (1+\sigma_{\ell}^z)/2$, and $\Delta
\to \lambda$.

Before ending this section, we note that the periodic dynamics of
$H_0$ has also been studied recently using a square pulse protocol
which drives $\lambda(t)$ between $\lambda$ and $-\lambda$ in the
regime $w/\lambda \ll 1$ \cite{bm1}. In particular, the stroboscopic
evolution of $O_{22}$ as a function of the drive cycle $n$ for
several frequencies starting from the $|\mathbb{Z}_2\rangle$ state
has been shown to display long-term coherent oscillations in the
high drive frequency regime. This behavior has been tied to the
presence of scars in the Floquet Hamiltonian of the driven system.
At low frequencies, scars were absent and the system displayed
thermalization consistent with ETH. In between, at moderate drive
frequencies, $O_{22}$ shows several reentrant transitions between
thermal and coherent behavior. In what follows, we are going to
perform a similar study for $H_0$ in the presence of random and
quasiperiodic drive protocols.

\section{Random drive protocol}
\label{sec3}

In this section, we shall address the dynamics of the system
described by $H_s$ (Eq.\ \ref{spinham1}) in the presence of a random
sequence of square pulses which makes the parameter $\lambda$ time
dependent. In this work, we shall be interested in the regime where
$w \ll |\lambda(t)|$ throughout the drive cycle. The randomness
corresponds to a telegraphic noise in the drive protocol leading to
a time period of $T_{\pm} = T+ \alpha dT$, where $\alpha= \pm 1$ is
a random number and $dT$ denotes the strength of the noise. Under
such a drive $\lambda(t) = +(-) \lambda$ for $t>(\le) T_{\pm}/2$.

To understand the effect of such a random drive, we first note that
in the absence of randomness ($dT=0$), the dynamics of $H_s$, for $w
\ll \lambda$, has been studied in Ref.\ \onlinecite{bm1}. It was
found that to ${\rm O}(w/\lambda)$, the Floquet Hamiltonian of the
system is given by
\begin{eqnarray}
H_F = -w \frac{\sin (\gamma)}{\gamma} \sum_j \left( \cos (\gamma) \tilde
\sigma_j^x + \sin(\gamma) \tilde \sigma_j^y \right) + ... \label{f1}
\end{eqnarray}
where $\gamma= \lambda T/(4 \hbar)$, the ellipsis corresponds to
${\rm O}(w^3)$ and higher order terms whose analytical form is
unknown, and $T= 2\pi/\omega_D$ is the drive period. The
corresponding unitary evolution operator is given by $U= \exp[-i H_F
T/\hbar]$. It was found that the ${\rm O}(w)$ term constitutes a
renormalized PXP Hamiltonian \cite{bm1} which vanishes for $\gamma=
n\pi$. At these points, the Floquet Hamiltonian consists of ${\rm
O}(w^3)$ (and higher powers of $w$) terms which have different
structure compared to PXP Hamiltonian. For $\gamma \ne n\pi$, the
${\rm O}(w)$ term has the most dominant contribution in $H_F$.
Moreover, the form of these ${\rm O}(w)$ terms ensures that when
they are dominant, the Floquet spectrum hosts scars which lead to
long-time coherent dynamics. In contrast, for $\gamma= n \pi$,
eigenstates of the Floquet Hamiltonian do not host scars and the
system exhibits thermalization consistent with ETH.

For random drives it is easy to see that for $T= T_{\pm}$, the
unitary evolution operators controlling the evolution are given by
\begin{eqnarray}
U_{\pm} &=& e^{- i H_F^{\pm} T_{\pm}}, \quad H_F^{\pm} = H_F(\gamma
\to \gamma_{\pm}) \label{ranevol1}
\end{eqnarray}
where $\gamma_{\pm} = \lambda T_{\pm}/(4 \hbar) = \gamma \pm
d\gamma$, and $d \gamma = \lambda dT/(4\hbar)$. Thus for a random
protocol, the wavefunction after $n$ cycles of the drive would be
\begin{eqnarray}
|\psi_n\rangle = U_- U_- U_+ ... U_+ |\psi_0\rangle = {\mathcal
U}|\psi_0\rangle  \label{ranevol2}
\end{eqnarray}
where $|\psi_n \rangle$ denotes the wavefunction after $n$ drive
cycles starting from the initial state $|\psi_0\rangle$, and $U_+$
and $U_-$ occurs randomly with equal probability in the string of
evolution operators represented by ${\mathcal U}$ in Eq.\
\ref{ranevol2}.

In the presence of such a drive, the effect of randomness manifests
itself through the action of the commutator $[U_+,U_-]$ on the
state. This can be easily seen by noting that the Floquet
eigenvectors corresponding to $U_{\pm}$ changes only when it is
operated on by a subsequent $U_{\mp}$ in the random string in Eq.\
\ref{ranevol2}. This change occurs since eigenvectors of $U_+$ and
$U_-$ are different; it vanishes if $U_+$ and $U_-$ commute. Such
commutation of $U_+$ and $U_-$ clearly occurs for $dT=0$ since it
amounts to absence of randomness. However, in the $w \ll
|\lambda(t)|$ limit, the leading terms of these commutators also
vanish at special values of $dT/T$. To see this we compute
${\mathcal C}= [U_+,U_-]$. Using Eqs.\ \ref{f1} and \ref{ranevol1}
we find
\begin{eqnarray}
{\mathcal C} &=& {\mathcal C_0}+... \nonumber\\
&=&\left(\frac{4w}{\lambda}\right)^2 \sin (2 d\gamma)
\sin(\gamma+d\gamma) \sin(\gamma-d\gamma) \nonumber\\
&& \times \sum_{j,j'} \left[ \tilde \sigma_j^y, \tilde \sigma_{j'}^x
\right] + ... \label{ranevol3}
\end{eqnarray}
where the expression is valid for $w/\lambda \ll 1$ and the ellipsis
beyond $\mathcal{C}_0$ indicate higher order terms in $w/\lambda$.
We note that if the norm of the commutator vanishes, it is possible
to rearrange $U_{-}$ and $U_+$ in Eq.\ \ref{ranevol2} in pairs.
Since in the random string of evolution operators in Eq.\
\ref{ranevol2}, the occurrence of $U_+$ and $U_-$ are equally
likely, for large enough $n$, the dynamics could have been described
by an average Floquet Hamiltonian:  $|\psi(t)\rangle \simeq \exp[-i
H_F^{\rm av} nT] |\psi_0\rangle$, where
\begin{eqnarray}
H_F^{\rm av} &=& \left(H_+ T_+ + H_-T_-\right)/(2T) +... \nonumber\\
&=& \frac{w}{\gamma} \sum_j \left[c_1(T) \tilde \sigma_j^x + c_2(T)
\tilde \sigma_j^y \right] +... \nonumber\\
c_1(T) &=& \sin(2[\gamma + d\gamma]) + \sin(2[\gamma-d\gamma]) \nonumber\\
c_2(T) &=& 2-\cos(2[\gamma + d\gamma]) -\cos(2[\gamma-d\gamma])
\label{randevol4}
\end{eqnarray}
where ellipsis indicate terms ${\rm O}(w^m)$ for $m \ge 3$ which do
not support scars \cite{bm1}.

\begin{figure}
\rotatebox{0}{\includegraphics*[width=0.49 \linewidth]{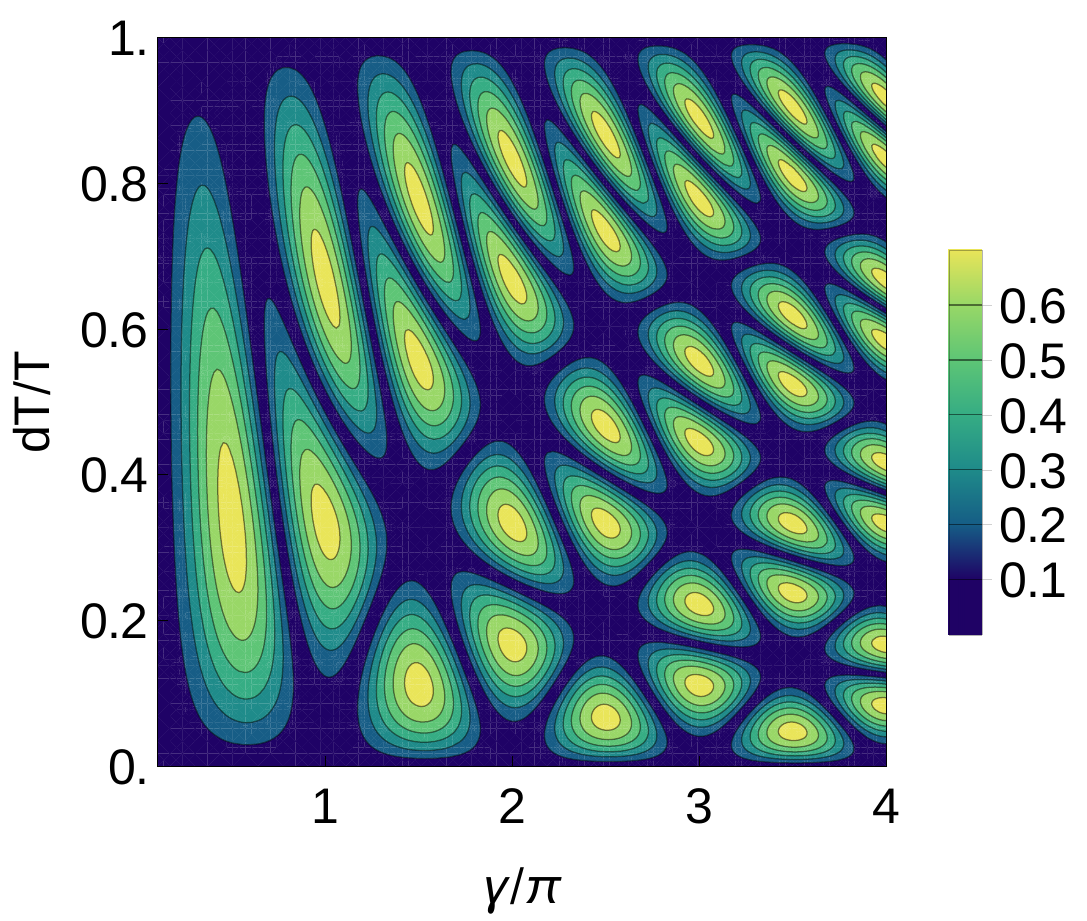}}
{\includegraphics*[width=0.49 \linewidth] {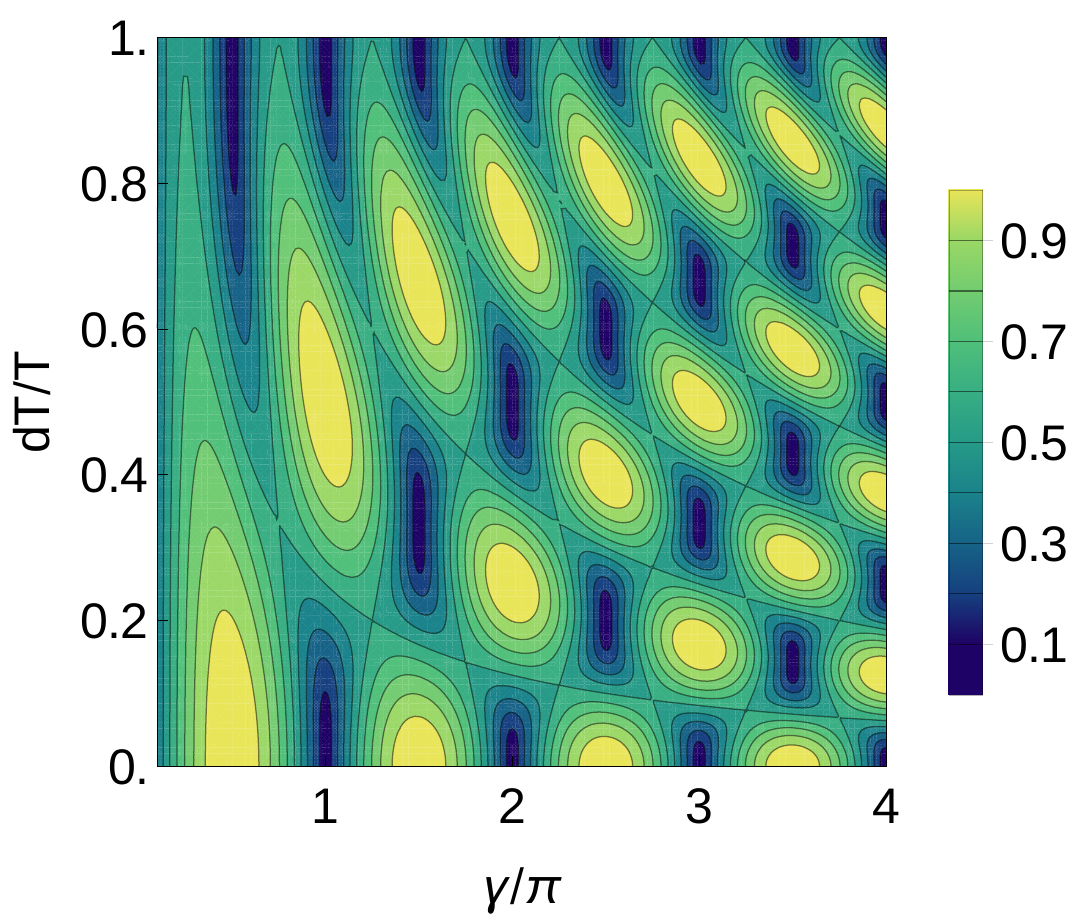}} \caption{
Left Panel: Plot of $||{\mathcal C}_0||(\lambda/4w)^2$ as a function
of $\gamma$ and $dT/T$ showing dips at $dT/T=k\pi/2$ and
$k\pi+\gamma$. Right: Plot of $||H_{F}^{\rm av}||/w$ as a function
of $\gamma$ and $dT/T$. \label{fig1}}
\end{figure}

\begin{figure}
\rotatebox{0}{\includegraphics*[width=\hsize]{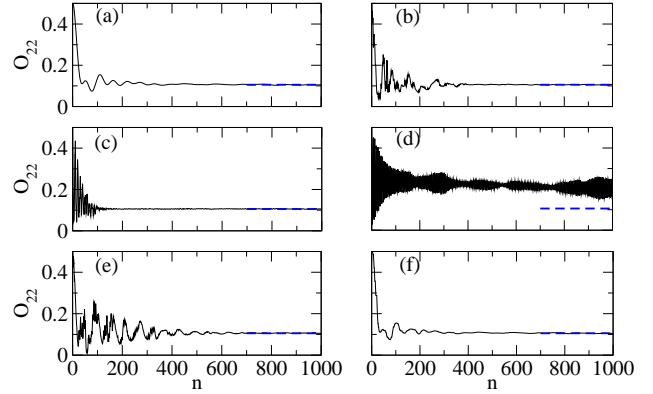}}
\caption{ Plot of $O_{22}$ as a function of the number of drive
cycle $n$ for $L=26$, $\omega_D=7.75$ and $\lambda=15$ for (a)
$dT/T=0$, (b) $0.1$, (c) $0.3$, (d) $0.5$, (e) $0.9$, and (f) $1$.
The plot indicates clear return of coherent oscillation for
$dT/T=0.5$. All energies (frequencies) are scaled in units of $w/\sqrt{2}$
($w/(\hbar \sqrt{2})$) and $\hbar$ is set to unity. The blue dashed
line in all panels corresponds to the infinite temperature ensemble
value of $O_{22}$.} \label{fig2}
\end{figure}

Using Eqs.\ \ref{ranevol3} and \ref{randevol4}, we can now chart out
analytical conditions for having long-time coherent oscillations in
the presence of the random drive. The first condition for such
oscillation is sufficiently weak randomization of Floquet
eigenstates which occurs when the leading term in norm of commutator
$||{\mathcal C}_0||= (4w/\lambda)^2\sin (2 d\gamma)
\sin(\gamma+d\gamma) \sin(\gamma-d\gamma)$ vanishes. This leads to
the condition
\begin{eqnarray}
d\gamma &=& k \frac{\pi}{2} \quad {\rm or} \quad d\gamma = k\pi \pm
\gamma  \label{randevolcond1}
\end{eqnarray}
where $k \in \mathbb{Z}$. We note that Eq.\ \ref{randevolcond1} constitutes a
necessary but not sufficient condition for coherent oscillations.
For such oscillations, in addition to weak enough randomization of
Floquet eigenvectors, one also needs to ensure that $H_F^{\rm av}$
which controls the dynamics when $||{\mathcal C}_0||=0$, hosts
scars. This requires an additional constraint that the leading term
in the norm of $H_{F}^{\rm av}$, $||H_{F}^{\rm av}||=w
\sqrt{c_1(T)^2 +c_2(T)^2}$, does not vanish ({\it i.e.}, $c_1(T)$ and
$c_2(T)$ do not vanish simultaneously). This leads to the
condition
\begin{eqnarray}
d \gamma &\ne& k' \pi \quad {\rm if} \quad \gamma = k \pi
\label{randevolcond2}
\end{eqnarray}
where $k, k' \in \mathbb{Z}$. We note that the conditions Eq.\
\ref{randevolcond1} and \ref{randevolcond2} ensure that the effect
of the telegraphic noise is minimal and that the dynamics is
controlled by scars in the eigenspectrum of $H_F^{\rm av}$. Thus
these points in the parameter space of the system is likely to host
coherent oscillations. These conditions are represented in Fig.\
\ref{fig1}. The left panel shows the regions in $dT/T-\gamma$ plane
where $||{\mathcal C}_0|| = 0$, while the right panel indicates
regions where $||H_{F}^{\rm av}||$ remains finite. The common
points between these two regions that satisfy both these criteria
are the ones where one expects restoration of coherence in the
presence of noise.

\begin{figure}
\rotatebox{0}{\includegraphics*[width=0.98 \linewidth]{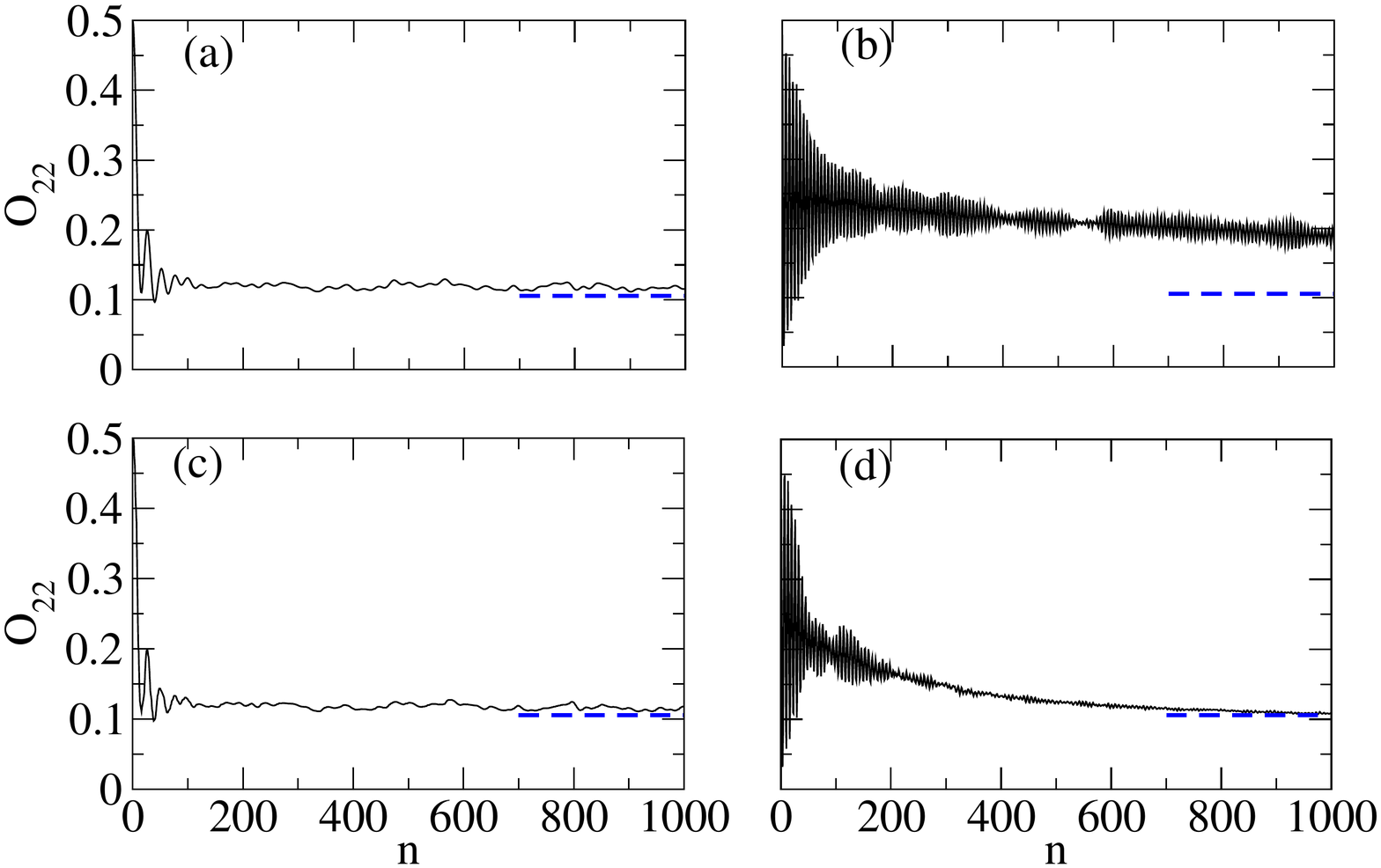}}
\caption{  Plot of $O_{22}$ as a function of $n$ for $L=26$,
$\omega_D=3.9$ and $\lambda=15$ (a) $dT/T=0$, (b) $0.25$, (c)$0.5$,
and (d) $0.75$. The coherent oscillatory behavior exhibits a much
slower decay time for $dT/T=0.25$ than for $dT/T=0.75$. All units
and the definition of the blue dashed line
are same as in Fig.~\ref{fig2}.} \label{fig3}
\end{figure}

To verify the restoration of coherence, we numerically compute
$O_{\ell 2}$. To this end, we note that both $U_+$ and $U_-$ can be
expressed in terms of the eigenvalues $\epsilon_p^{(1)[(2)]}$ and
eigenvectors $|p^{(1),[(2)]}\rangle$  of $H_s^{1[2]} = H_F(+[-]
\lambda)$ as
\begin{eqnarray}
U_{\pm}&=& \sum_{p,q} e^{-i(\epsilon_q^{(1)} +
\epsilon_p^{(2)})T_{\pm}/2} \mu_{pq}^{12} |p^{(2)}\rangle \langle
q^{(1)}| \label{uevoln1}
\end{eqnarray}
where $\mu_{pq}^{12} = \langle p^{(2)}|q^{(1)}\rangle$. These
eigenvalues and eigenvectors are computed numerically using
exact-diagonalization for finite-size system with $L\le 26$. Using
this, one can numerically compute $O_{\ell 2}$ as
\begin{eqnarray}
O_{\ell 2} &=& \langle \psi_0|{\mathcal U}^{\dagger} \sigma_{\ell}^z
\sigma_{\ell +2}^z {\mathcal U} |\psi_0 \rangle \label{numeval1}
\end{eqnarray}
We note that translational symmetry ensures that $O_{\ell 2}$ is
independent of $\ell$; in what follows we shall therefore
concentrate on $O_{22}$.

\begin{figure}
\rotatebox{0}{\includegraphics*[width=0.98 \linewidth]{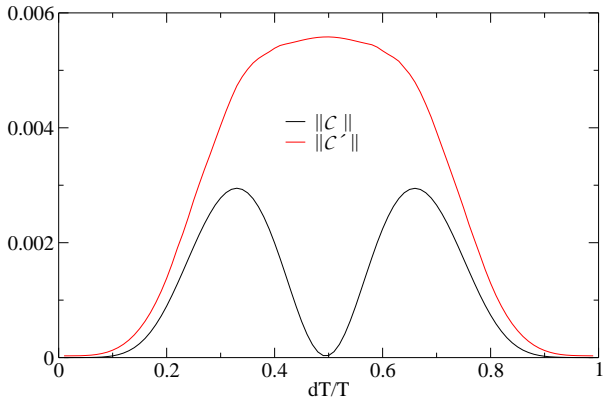}}
\caption{ Plot $||{\mathcal C}||$(black solid line) and $||{\mathcal
C}'||$ (red solid line) (both normalized by the square of the
Hilbert space dimension)for $L=26$ and $\omega_D=7.75$ as a function
of $dT/T$ showing clear dip of $||{\mathcal C}||$ at $dT/T=1/2$. All units
are same as in Fig.~\ref{fig2}.} \label{fig4}
\end{figure}
The numerical plot of $O_{22}$, shown in Figs.\ \ref{fig2} and
\ref{fig3}, supports the expectation obtained from the analytical
consideration charted out earlier in the section. In Fig.\
\ref{fig2}, we plot $O_{22}$ as a function of the number of drive
cycles $n$, for $\omega_D=7.75$ (which corresponds to $\gamma \simeq
2$) and for several representative values of $dT/T$. We note that
for these systems, for $dT=0$, the system shows rapid thermalization
consistent with ETH as pointed out in Ref.\ \onlinecite{bm1}. Here
we find that coherent oscillations pick as we tune towards
$dT/T=0.5$, where $||{\mathcal C_0}||=0$ and $||H_{F}^{\rm av}||
\ne 0$. At $dT/T =0.5$, the system exhibits long-time coherent
oscillations and constitutes an example of coherence restoration by
temporal disorder. To check that this is indeed the case, we plot
$||{\mathcal C}|| = ||\left[U_+,U_-\right]||$ and $||{\mathcal C}'||
=||U_+ U_-||$ (normalized by the square of the Hilbert space
dimension) for $\omega_D=7.75$ as a function of $dT/T$. The plot,
shown in Fig.\ \ref{fig4}, indicates a clear dip of $||{\mathcal
C}||$ at $dT/T=0.5$ where $||{\mathcal C}'||$ remain finite. This
corroborates our expectation from the earlier discussion based on
analysis of $||{\mathcal C}_0||$ and $||H_{F}^{\rm av}||$.

\begin{figure}
\rotatebox{0}{\includegraphics*[width=0.98 \linewidth]{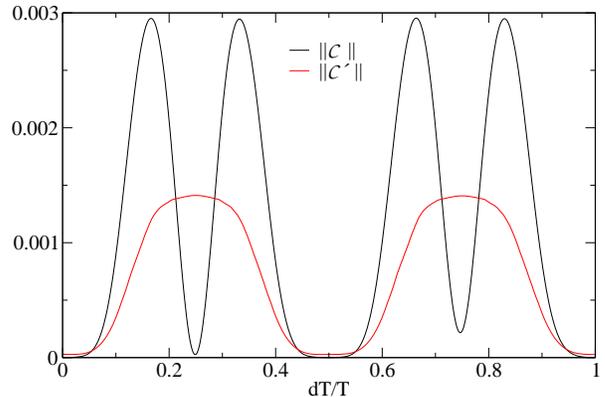}}
\caption{Plot of $||{\mathcal C}||$(black solid line) and
$||{\mathcal C}'||$ (red solid line) (both normalized by the square
of the Hilbert space dimension) for $L=26$ and $\omega_D=3.9 $ as a
function of $dT/T$ showing clear dip of $||{\mathcal C}||$ at
$dT/T=1/4, \, 1/2, \, {\rm and}\, 3/4$. Note that $||{\mathcal
C}'||$ also vanishes around $dT/T=1/2$ whereas it remains finite for
$dT/T=1/4\, {\rm and}\, 3/4$.  All units are same as in Fig.~\ref{fig2}.} \label{fig5}
\end{figure}
A similar noise-induced restoration of coherence is seen in Fig.\
\ref{fig3} for $\omega_D=3.9$ ($\gamma \simeq 4$) where long-time
coherent oscillations of $O_{22}$ returns at $dT/T=1/4,3/4$. This is
in accordance with prediction of Eqs.\ \ref{randevolcond1} and
\ref{randevolcond2}. Note that in this case $||{\mathcal C}_0||=0$
at $dT/T=0.5$; however $||H_{F}^{\rm av}||$ also vanishes at this
point and $O_{22}$ does not exhibit long-time oscillations. The
analytical prediction is further verified by numerical plot of
$||{\mathcal C}||$ and $||{\mathcal C}'||$ at $\omega_D=3.9$ as a
function of $dT/T$ as shown in Fig.\ \ref{fig5}. We find clear dip
in $||{\mathcal C}||$ at $dT/T=1/4,\, 1/2,\, 3/4$. However, at
$dT/T=1/2$, $||{\mathcal C}'||$ also vanishes leading to absence of
long-term coherent oscillations as discussed. Finally, we note that
the restoration of coherence is more robust at $dT/T=1/4$ compared
to $dT/T=3/4$. This feature can be qualitatively understood as
follows. We first note that the thermalization in these systems
leading to destruction of coherence occurs due to action of
${\mathcal C}$; thus $||{\mathcal C}||$ is an indicator of the
strength of this term. Next, we note that such terms lead to finite
matrix elements between states within the scar subspace and states
within the ETH band. This can be checked by noting that ${\mathcal
C}_0$ (the first term in Eq.\ \ref{ranevol3}) indeed leads to such
matrix elements. Thus it is expected that the thermalization time of
$O_{22}$, $\tau_{\rm th}$, would depends on $||{\mathcal C}||$. The
expression of $\tau_{\rm th}$ can be estimated using Fermi's golden
rule and assuming a constant density of state $\rho_0$ for states in
the thermal band to be $\tau_{\rm th}^{-1} \simeq (2 \pi/\hbar)
\rho_0 ||{\mathcal C}||^2 \sim ||{\mathcal C}||^2$. Thus a larger
$||{\mathcal C}||$ is expected to lead to shorter thermalization
time and faster loss of coherence. This features is manifested in
relatively shorter thermalization time of oscillations for
$dT/T=0.75$ in Fig.\ \ref{fig3} compared to those for $dT/T=0.25$.
Although $||{\mathcal C}_0||$ vanishes in both case, the remaining
terms lead to a larger $||{\mathcal C}||$ and hence shorter
thermalization time for $dT/T=0.75$.

\section{Quasiperiodic drive protocol}
\label{sec4}

In this section, we will study the dynamics of the system when it is driven by
a Thue-Morse sequence (TMS) generated by the two evolution operators $U_+$ and
$U_-$ given in Eq.~\eqref{ranevol1}. The motivation for this is that a TMS
generated by two
non-commuting operators is known to generate unusual long-time behaviors which
are quite different from those generated by a random sequence~\cite{tmdyn1}.
The TMS is generated as follows~\cite{tmdyn1}. Defining $A_0 = U_+$ and
$B_0 = U_-$, we recursively define
\begin{eqnarray} A_{m+1} &=& B_m A_m, \quad
B_{m+1} = A_m B_m, \label{tms1} \end{eqnarray} for all $m \ge 0$.
The wave function after $2^n$ drives is then given by
\begin{equation} | \psi_{2^n} \rangle ~=~ A_n | \psi_0 \rangle.
\label{tmsn} \end{equation}
For instance, the wave function after $2^3 = 8$ drives is
\begin{equation} | \psi_8 \rangle ~=~ A_3 | \psi_0 \rangle ~=~ U_- U_+ U_+ U_-
U_+ U_- U_- U_+ | \psi_0 \rangle. \label{tms8} \end{equation}
It is clear that one can use the recursion relations in Eq.~\eqref{tms1} to
generate a sequence of $2^n$ drives by performing only $2n$ matrix
multiplications. This enables us to study relatively easily what happens after
an exponentially large number of drives.

\begin{figure}
\includegraphics*[width=0.98\linewidth]{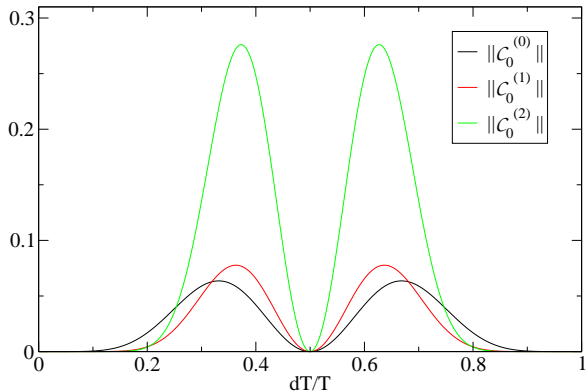}
\caption{Plots of $||{\mathcal C}_0^{(0)}||$ (black line), $||{\mathcal C}_0^{(1)}||$ (red line) and $||{\mathcal
C}_0^{(2)}||$ (green line) versus $dT/T$, for $\omega_D = 7.5$, and $\lambda = 15$.  All units
are same as in Fig.~\ref{fig2}.} \label{fig6}
\end{figure}

We will now study the conditions under which the dynamics generated
by Eqs.~(\ref{tms1}-\ref{tmsn}) gives long-time coherent
oscillations. As discussed in Sec.~\ref{sec3}, this will happen if
the evolution operators $A_0$ and $B_0$ commute, but $A_1 = B_0 A_0$
is not equal to the identity operator and it has scar states as its
eigenstates. However, the recursive form of Eq.~\eqref{tms1} implies
that even if $A_0$ and $B_0$ do not commute, we can still obtain
long-time coherent oscillations if $A_1$ and $B_1$ commute (since
the unitary dynamics after an even number of drives can be written
solely in terms of $A_1$ and $B_1$), but $A_2 = B_1 A_1$ is not
equal to the identity and it hosts scar states. Clearly, this idea
can work at higher and higher levels. We thus obtain a hierarchy of
possibilities for getting coherent oscillations, given by the
condition that although $A_{n-1}$ and $B_{n-1}$ do not commute,
$A_n$ and $B_n$ commute and $A_{n+1} = B_n A_n$ is not equal to the
identity and it hosts scars. We define the norm of the level $n$
commutator as $||{\cal C}^{(n)}|| \equiv || [A_n, B_n] ||$. To
demonstrate this point, we consider the ${\rm O}(w)$ approximation
to $H_F$ and write $U_{\pm} = \exp[-i H_{F}^{\pm}T/\hbar]$, where
$H_F^{\pm}$ is given by Eq.\ \ref{f1} with $\gamma \to
\gamma_{\pm}$. Using these we construct the matrices $A_1 = U_+ U_-$
and $B_1 = U_- U_+$ and compute $||{\mathcal C}_0^{(1)}||$. A
similar procedure leads to $||{\mathcal C}_0^{(2)}||$. Here, the
subscript $0$ in $\mathcal{C}_0^{(n)}$ refers to the fact that only
the ${\rm O}(w)$ approximation to $H_F$ was used for the
computations. Fig.~\ref{fig6} shows plots of $||{\cal C}_0^{(n)}||$
versus $dT/T$ for $n=0$, $1$ and $2$, when $\omega = 7.5$,
$w=1/\sqrt{2}$, and $\lambda = 15$. These norms are seen to approach
zero for a range $dT/T \le 0.2$ and $dT/T \ge 0.8$
implying that coherent oscillations can be expected to occur around
such values. We note however that these expectations from a ${\rm
O}(w)$ theory is qualitative; clearly higher order terms are
expected to reduce this range to (possibly) discrete points. A more
precise investigation of this behavior requires exact numerics which
we now carry out.

\begin{figure}
\rotatebox{0}{\includegraphics*[width=0.98 \linewidth]{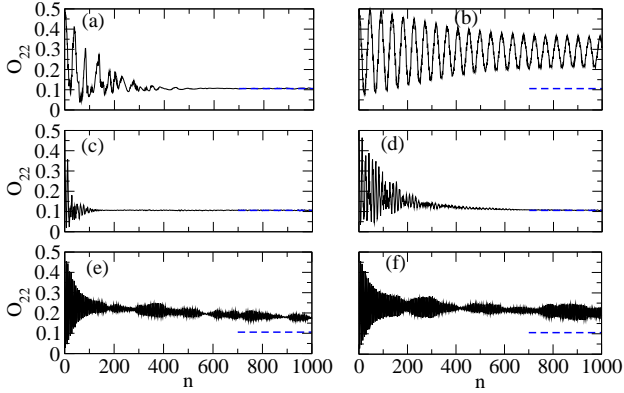}}
\caption{Comparison of $O_{22}$ as a function of $n$ for random
(left column) and TMS (right column) at some special points, with
$L=26$, $\omega_D=7.5$, and $\lambda=15$. (a)-(b) $dT/T=0.1$, (c)-(d)
$dT/T=0.25$, and (e)-(f) $dT/T=0.5$.  All units
and the definition of the blue dashed line
are same as in Fig.~\ref{fig2}. } \label{fig7}
\end{figure}

The plot of the correlation function $O_{22}$ as a function of $n$
for $\omega_D=7.5$, $w=1$, $\lambda=15$ and several representative
values of  $dT/T$ is shown in Fig.\ \ref{fig7}. The left panel shows
results for the random protocol while the right panel shows that for
TMS. We note that similar to the random sequence discussed in
Sec.~\ref{sec3}, the TMS also leads to quick thermalization when
$||{\mathcal C}_0^{(0)}||\neq 0$, and to coherent oscillation when
$||{\mathcal C}_0^{(0)}||=0$ but the leading term in $||H_{F}^{\rm
av}||/w$ is non-zero. This can be from Fig.~\ref{fig7}(e) and (f)
where both the random (Fig.\ \ref{fig7}(e)) and TMS (Fig.\
\ref{fig7}(f)) at $dT/T=0.5$ display coherent oscillations. However,
as shown in the top panel ($dT/T=0.1$) of Fig.\ \ref{fig7}, TMS may
lead to oscillatory behavior at special values of $dT/T$ (Fig.\
\ref{fig7}(b)) even when random protocol leads to thermalization
(Fig.\ \ref{fig7}(c)). In the middle panel of Fig.\ \ref{fig7}, a
comparison between the behavior of $O_{22}$ driven by random (Fig.\
\ref{fig7}(c)) and TMS (Fig.\ \ref{fig7}(d)) also indicates a much
longer thermalization time for the latter. This can be understood to
be a precursor to the oscillatory behavior of $O_{22}$ for TMS at
$dT/T=0.3$, analogous to that found for $dT/T=0.1$.

\begin{figure}
\rotatebox{0}{\includegraphics*[width=0.98 \linewidth]{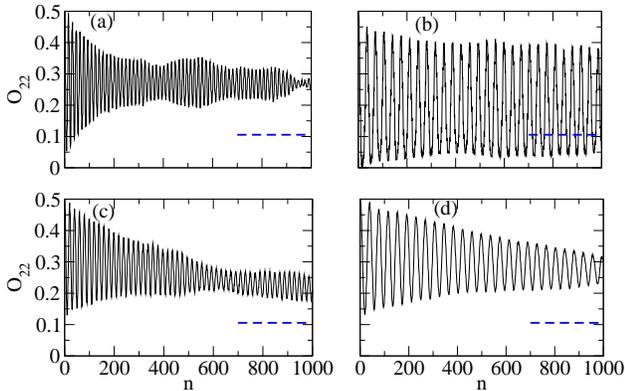}}
\caption{Plot of $O_{22}$ as a function of $n$ for the TMS showing
oscillatory behavior for (a) $dT/T=0.1$, $\omega_D=6.75$ (b) $dT/T=0.1$, $\omega_D=8.25$ (c) $dT/T=0.3$, $\omega_D=2.5$, and (d)
$dT/T=0.3$, $\omega_D=5$. For all these parameter values, the random protocol
shows rapid thermalization. All units
and the definition of the blue dashed line
are same as in Fig.~\ref{fig2}.} \label{fig8}
\end{figure}

It turns out there are several such special points in the ($T$,
$dT$) parameter space where the random and TMS show drastically
different behaviors, namely, rapid thermalization for the random
sequence but coherent oscillations for the TMS. We demonstrate four
other such point in Fig.~\ref{fig8}. For all such points, the random
drive leads to rapid thermalization. The coherent behavior of
$O_{22}$ thus reflects the quasiperiodic nature of the TMS which is
distinct from a totally random sequence. At these parameter values,
the special form of the noise correlation in the TMS (i.e., the
particular form of the sequence of $U_+$'s and $U_-$'s), although
not comparable to a perfectly periodic sequence, is sufficient to
preserve the memory of the initial $|\mathbb{Z}_2\rangle$ state for
a long time.

\begin{widetext}

\begin{figure}
\includegraphics[width=\linewidth]{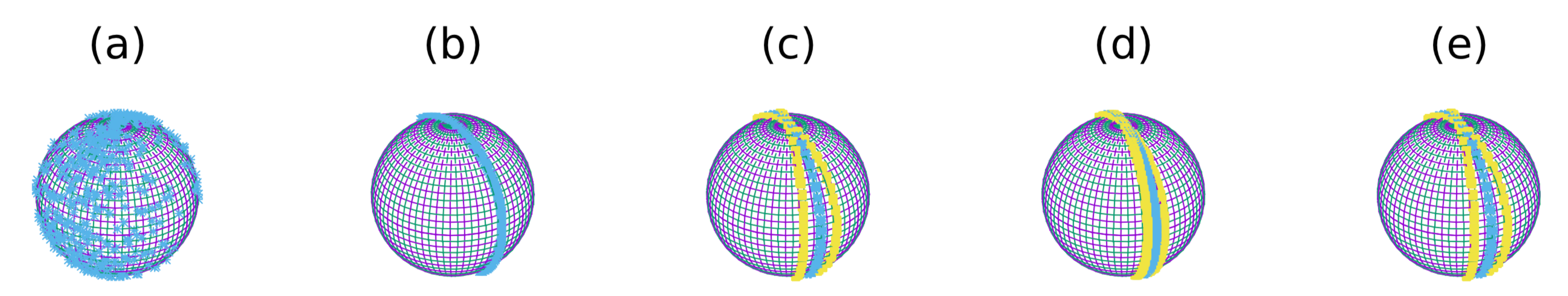}
\caption{ Motion of $\psi_n$ on Bloch sphere for $\lambda=15$, and ($\omega_D$,
$dT/T$) = (a) (7.5, 0.3), (b) (7.5, 0.5), (c) (7.5, 0.1), (d) (5.0, 0.3), (e)
(2.5, 0.3). In (c)-(e) the yellow (blue) circles are for odd (even) values of
$n$. See text for details.  All units are same as in Fig.~\ref{fig2}.} \label{fig9} \end{figure}

\end{widetext}

\begin{figure}
\includegraphics[width=\linewidth]{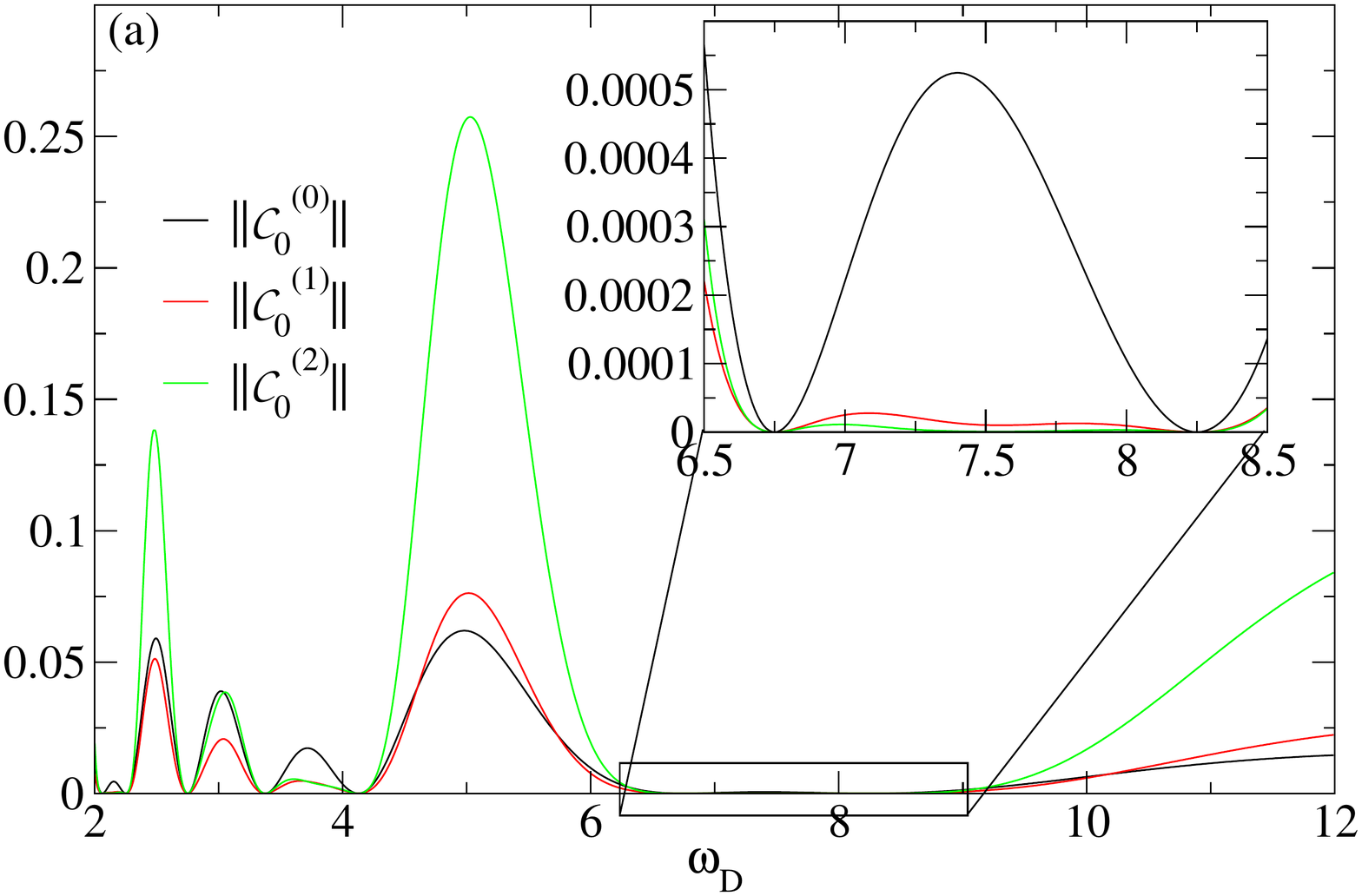}\\
\includegraphics[width=\linewidth]{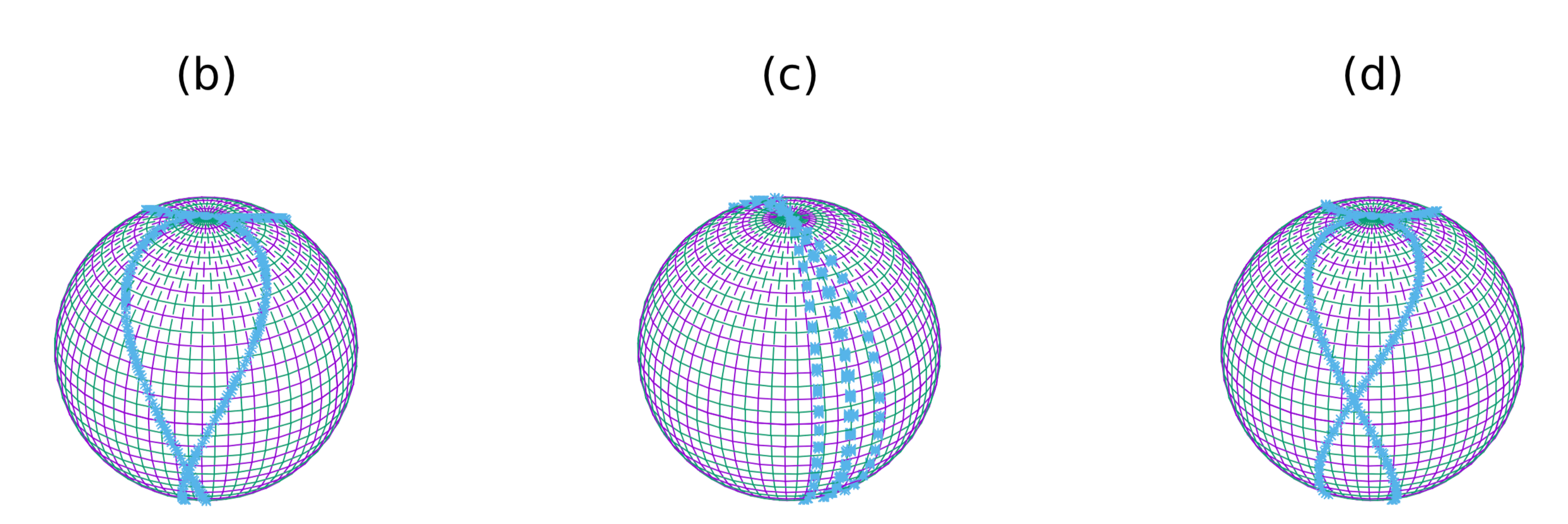}
\caption{ (a) Plots of $||{\mathcal C}_0^{(0)}||$ (black line),
$||{\mathcal C}_0^{(1)}||$ (red line) and $||{\mathcal C}_0^{(2)}||$
(green line) versus $\omega_D$ for a fixed $dT/T=0.1$, and $\lambda
= 15$. Motion of $\psi_n$ on Bloch sphere for $\lambda=15$,
$dT/T=0.1$ and $\omega_D$ equal to (b) 6.75 (c) 7.5 and (d) 8.25.
All units are same as in Fig.~\ref{fig2}.} \label{fig10}
\end{figure}

Due to the aperiodic nature of the TMS, it seems difficult to
describe the special points based on an effective many-body Floquet
Hamiltonian. However, we find that it is possible to find the
positions of these special points without studying the exact
many-body dynamics (which is numerically difficult). We demonstrate
this by studying the dynamics of a two-level system governed by the
following Hamiltonian
\begin{eqnarray}
H_{2 \times 2} ~= -w ~\frac{\sin \gamma}{\gamma} \left( \cos \gamma
~ \sigma^x + \sin \gamma ~\sigma^y \right) . \label{2x2}
\end{eqnarray} This is basically the ${\rm O}(w)$
Hamiltonian in Eq.~\eqref{f1} but with only one site. We now
calculate $U_{\pm}$ using this Hamiltonian and the driven wave
function by acting with $U_+$ and $U_-$ on the initial state
($\psi_0=(1,0)^T$) for a total of $n_{tot}$ times following the TMS
(we choose $n_{tot}=1000$). The two-component driven wave function
$\psi_n$ can be mapped to the Bloch sphere ($\theta(n), \phi(n)$)
using the parametrization
\begin{eqnarray}
\psi(n)=(\cos(\theta/2)e^{i\phi/2}, \sin(\theta/2)e^{-i\phi/2})^T.
\label{paramet1}
\end{eqnarray}
We find completely chaotic motion of $\psi_n$ on the Bloch sphere
for parameters which show quick thermalization to the infinite
temperature ensemble in the exact many-body dynamics (see
Fig.~\ref{fig9} (a)). On the other hand, $\psi_n$ follows a regular
trajectory on the Bloch sphere when we have coherent oscillations in
the exact many-body dynamics. We find that this coherent behavior
can be further categorized into {\it at least} three classes. For
parameter values where $||{\mathcal C}_0^{(0)}||=0$ but the leading
term in $||H_{F}^{\rm av}||/w$ is non-zero, the trajectory is just a
single circle (see Fig.~\ref{fig9} (b)), whereas at the special
points, we see either three circles (see Figs.~\ref{fig9} (c)-(e) or
Fig.~\ref{fig10} (c)) or a closed curve made of intertwined ellipses
(see Fig.~\ref{fig10}(b),(d)). In fact, both the latter cases are
encountered when (for example) $dT/T$ is kept fixed at $0.1$ and
$\omega_D$ is varied (see Fig.~\ref{fig10} (a)). In the three circle
case, we further see that $\psi_n$ for even values of $n$ are
concentrated on one circle, while $\psi_n$ for odd values of $n$ are
concentrated on the other two circles. The one circle and the three
circle cases can be understood using the recursive structure for the
TMS. When $||{\mathcal C}^{(0)}_0||=0$, $U_+$ and $U_-$ commute with
each other; this implies that these can be written as
\begin{eqnarray} U_+ &=& e^{i \alpha_+ {\hat n}_+ \cdot {\vec \sigma}},
\quad U_- =  e^{i \alpha_- {\hat n}_- \cdot {\vec \sigma}},
\label{upm}
\end{eqnarray}
where the unit vectors ${\hat n}_+$ and ${\hat n}_-$ are identical,
and $\alpha_\pm$ are non-zero. Hence every term in the TMS sequence
has the form given by $\exp (i f_n {\hat n}_+ \cdot {\vec \sigma})$,
where $f_n$ is a number which depends on the number of $U_+$'s and
$U_-$'s which appear in the $n$-th term of the TMS. The trajectory
of $|\psi_n \rangle$ therefore lies on a single circle on the Bloch
sphere. At other special points like in Fig.~\ref{fig9} (c)-(e) and
Fig.~\ref{fig10} (c), $U_+$ and $U_-$ do not commute with each
other, but $U_+ U_-$ and $U_- U_+$ approximately
commute with each other, namely, $||{\mathcal C}^{(0)}_0|| \ne 0$
but $||{\mathcal C}_0^{(1)}|| \simeq 0$. This implies that $U_+ U_-$
and $U_- U_+$ can be written in forms similar to Eq.~\eqref{upm},
with identical unit vectors. Hence after any even number of drives
(which are given by products of a certain number of $U_+ U_-$ and
$U_- U_+$), we will get points which lie on a single circle on the
Bloch sphere. But after an odd number of drives, we will get a point
which corresponds to the single circle mentioned above multiplied by
either $U_+$ or $U_-$ depending on which of the two appears at the
last drive; these will give two different circles as $U_+$ and $U_-$
do not commute. Then, there are other special points where the
trajectory on the Bloch sphere is not composed of a single or a
three circle but a more complicated closed curve (Fig.~\ref{fig10}
(b) and Fig.~\ref{fig10} (d)).

\begin{figure}
\rotatebox{0}{\includegraphics*[width=0.98 \linewidth]{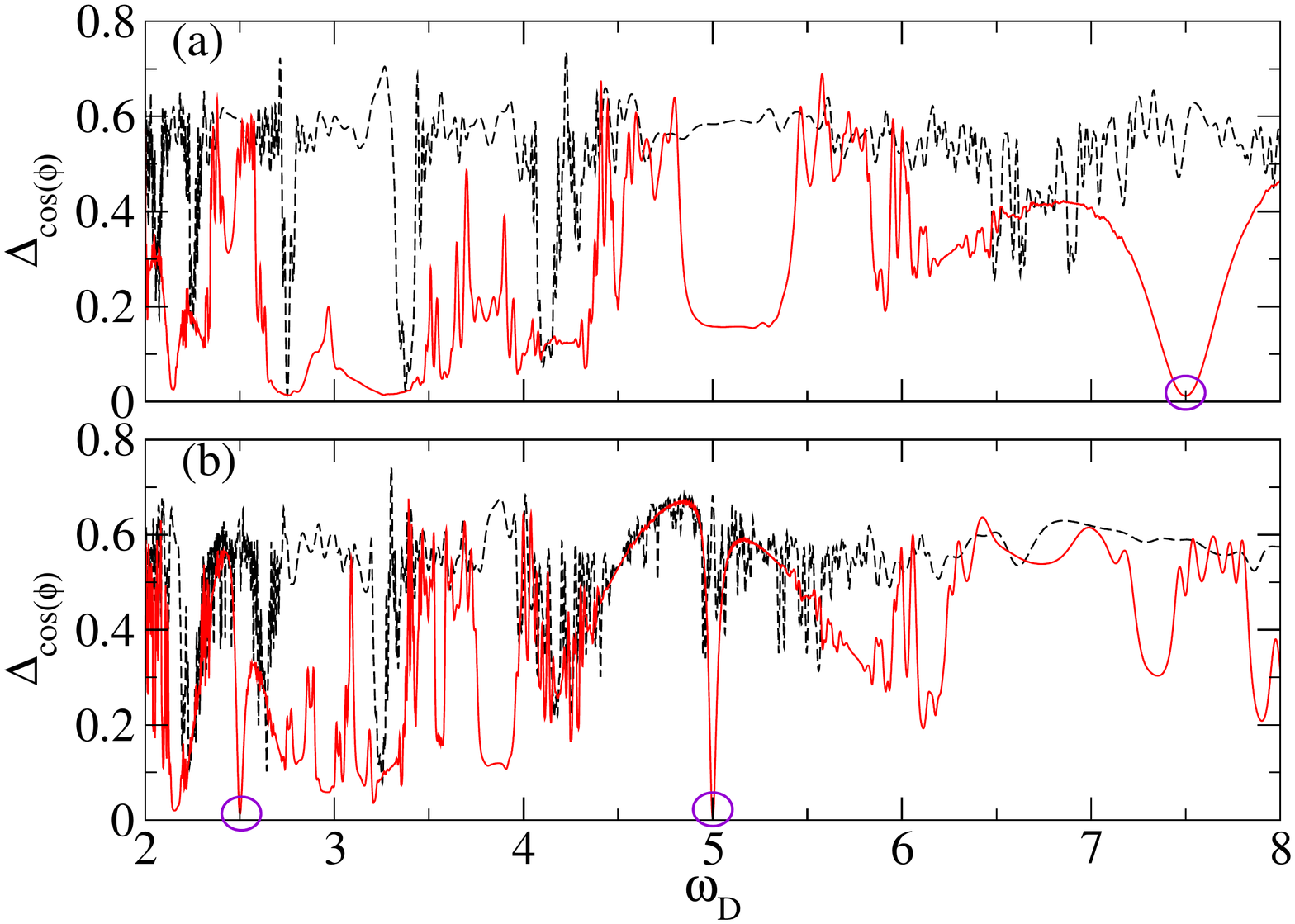}}
\caption{Plot of $\Delta_{\cos(\phi)}$ vs $\omega_D$ for $\lambda=15$,
$dT/T=$ (a) 0.1 (b) 0.3. Black dashed (red solid) lines denotes the graph
for random (TM) sequences. Violet circles denote the special points.
All units are same as in Fig.~\ref{fig2}.}
\label{fig11} \end{figure}

A single quantity as a function of $\omega_D$ would be useful to see
the rarity of the special points where coherent oscillations occur.
We note that a regular trajectory means that the fluctuation
$\Delta_{\cos(\phi)}$ in $\cos \phi(n)$ (where $\phi(n)$ denotes the
value of azimuthal angle $\phi$ after $n$ drive cycles) will be
small. We define
\begin{equation} \Delta_{\cos(\phi)} ~=~ \sqrt{\frac{\sum_{n=1}^{n_{tot}}
(\cos \phi(n)-\cos \phi_{av})^2}{n_{tot}}}, \end{equation} where
$\cos\phi_{av}= [\sum_{n=1}^{n_{tot}}\cos \phi(n)]/n_{tot}$. We plot
$\Delta_{\cos(\phi)}$ vs $\omega_D$ (for both random and TMS) and
mark the special points (characterized by prominent dips for only
the TMS) by violet circles in Fig.~\ref{fig11}.

The above observations suggest that the coherent oscillations at the
special points can be qualitatively understood based on a single
site problem; hence they are only due to the interplay between the
drive parameters ($T$, $dT$) and the drive sequence (TMS in this
case). Many-body effects hardly change the positions of these
special points.

\section{Discussion}
\label{sec5}

In this work, we have studied the driven tilted Bose-Hubbard model
for aperiodic drive protocols. Our results indicate that for both
random and quasiperiodic drives, the presence of aperiodicity can
lead to coherent behavior even when the system thermalizes in their
absence. We have presented an analytical, albeit qualitative,
explanation for this phenomenon and pointed out the role of quantum
scars behind it.

For random drive protocols, we find that there are specific points
in the $(T, dT)$ plane, where the commutator of the evolution
operators $U_{+} \equiv U(T+dT)$ and $U_- \equiv U(T-dT)$ vanish to
${\rm O} (w^2/\lambda^2)$. This means norm of such commutators become
extremely small at these points leading to minimal decoherence due
to noise. If at such points $U_{\pm}$ supports scars in their
Floquet Hamiltonian $H_F^{\pm}$ (note that while one cannot define
the Floquet Hamiltonian for the entire random string of $U_{+}$ and
$U_-$, each individual $U_+$ and $U_-$ have a well-defined $H_F$),
one sees coherent oscillations of correlation functions. We have
charted out the phase diagram in the $(T, dT)$ plane showing
existence and location of such points showing that random drives can
be instrumental in restoring coherence in an otherwise thermalizing
system which hosts quantum scars in its Floquet spectrum.

For the quasi-periodic drive protocol, we have chosen the Thue-Morse
sequence. We have shown that the inherent structure of such a drive
protocol leads to several additional coherence restoring points in
the $(T, dT)$ plane where the random protocol leads to thermalizing
behavior. We have plotted an approximate phase-space trajectory for
such drives on the local Bloch sphere using a simplified $2 \times
2$ local Hamiltonian. This analysis leads to four distinct class of
trajectories. Three of them, namely, chaotic, single circle and
three circles have a simple explanation as discussed here. However,
the intertwined elliptic trajectories does not seem to yield to a
simple qualitative explanation. We note here that similar
complicated dynamical behavior was studied for a single spin-$1/2$
subjected to a Fibonacci drive sequence in
Ref.~\onlinecite{Sutherland}. The generalization of this work to the
Thue-Morse sequence is left for future work.

The fluctuations of the azimuthal angle of
these trajectories are shown to provide a signature for coherent
behavior of the many-body system. It will be useful to understand why the
points at which coherence is restored in the full many-body driven
problem shifts so little from the results of this simplified
analysis. Furthermore, the mechanism and phase diagram of possible
coherence revivals using other forms of quasiperiodic drive
sequences, like the Fibonacci sequence~\cite{fib}, should also be
explored. We leave these issues as problems to be
explored in future works.

The model we have studied is known to provide a low-energy effective
description for ultracold Rydberg atoms on which quench experiments
have already been performed \cite{scarref1}. Here we suggest a drive
protocol where the detuning parameter is varied randomly with
periodicity $T+dT$ or $T-dT$. Our prediction, for example, is that
for starting from the regime $\Delta=15$ (in units of
$\sqrt{2}\Omega$) and $\omega_D=7.75$ (in units
$\sqrt{2}\Omega/\hbar$) where all values $dT < 0.5$ leads to rapid
thermalization, the Rydberg excitation density and density-density
correlation function will display long-time coherent oscillatory
behavior for $dT/T=0.5$. Richer, albeit similar, effects for
coherence restoration shall also be present for a quasiperiodic
(Thue-Morse) drive sequence as has been discussed here.

In conclusion, we have studied driven titled Bose-Hubbard model with
aperiodic drive. We have shown that the presence of randomness or
quasi-periodicity in the drive protocol may restore coherence in
such a driven system. We have provided analytic explanation of our
results, pointed out the role of quantum scars behind such coherent
behavior, and discussed the possibility of its experimental
signature in a driven ultracold Rydberg chain.

\begin{acknowledgements}
The work of A.S. is partly supported through the Partner Group
program between the Indian Association for the Cultivation of
Science (Kolkata) and the Max Planck Institute for the Physics of
Complex Systems (Dresden). D.S. thanks DST, India for Project No.
SR/S2/JCB-44/2010 for financial support.
\end{acknowledgements}

\end{document}